\documentclass[article]{aastex6}

\usepackage{CJK}

\begin{document}
\begin{CJK*}{UTF8}{bsmi}
\title{Identifying Rings in IFU Surveys}
\author{Chien-Hsiu Lee\altaffilmark{1}(李見修)}
\affil{Subaru Telescope, NAOJ, 650 North A'ohoku Place, Hilo, HI 96720, USA}

\altaffiltext{1}{leech@naoj.org}



\section{Introduction}
According to general relativity \citep{1915SPAW.......844E},
massive galaxies can induce strong
space-time curvature, gravitationally focus the lights of
background sources to form multiple images along the observer's
line-of-sight. When the lens, source, and the observer are
well aligned, the lensed image will become a ring, 
so-called Einstein ring \citep{1936Sci....84..506E}.
Despite their usefulness, only dozens of Einstein rings have been reported so far
\citep[see e.g.][]{2008ApJ...682..964B,2013MNRAS.436.1040S}, limited by the low resolution images from ground-based telescopes.
There were attempts to identify gravitational
arcs using imaging \citep[e.g.][]{2016A&A...592A..75P} or spectroscopy \citep{2008ApJ...682..964B}.
However, these searches require extensive modeling and fitting.
To minimize the efforts of finding heavily blended lens, we
propose to identify strong gravitational lensing candidates using IFU surveys, with the aid of computer vision techniques.

\section{Concept}
Because the space-time curvature is proportional to
the underlying masses, gravitational arcs are more likely to be found
around massive galaxies, i.e. luminous red galaxies (LRGs).
However, these luminous galaxies often outshine
the gravitational arcs. Indeed, it has been shown that there are numerous 
arcs heavily blended in the light profile of bright lenses,
not resolvable from the poor spatial resolution imaging obtained by ground-based telescopes.
One conventional method to identify heavily blended arcs is to search for
emission
features on top of the LRG spectrum. However, such approaches have only been
applied to long-slit or single fiber spectra; thus can only identify possible
candidates and still require high resolution imaging (e.g. from \textit{Hubble Space Telescope} or ground-based telescopes with adaptive optics)
to confirm the arcs. We propose to expand this approach to
IFU surveys and reveal arcs in the spatially resolved spectra.
Indeed, \cite{2017MNRAS.464L..46S} has analyzed IFU spectra from MaNGA, and
identified a strongly lensed system SDSSJ170124.01+372258.0 with dedicated modeling to remove the spectral
contributions from the LRG. However, it is time consuming to
model and remove LRG spectral contribution, hence \cite{2017MNRAS.464L..46S} 
only analyzed 81 galaxies with $\sigma>$250km/s. Alternatively, we can use
image subtraction to reveal arcs in the IFU data. As arcs have emission
features on top of the LRG spectra,
subtracting IFU images from subsequent wavelength will remove the LRG spectra and
reveal arcs. After the arcs are revealed, we can adopt computer vision
techniques, e.g. the Hough transform \citep{1962USP.3069654H,1972CACM..15..15D}, to
replace human inspection and automatically identify
the arc patterns, as demonstrated in \cite{2017PASA...34...14L}.

\section{Application}
\begin{figure*}
  \includegraphics[width=\columnwidth]{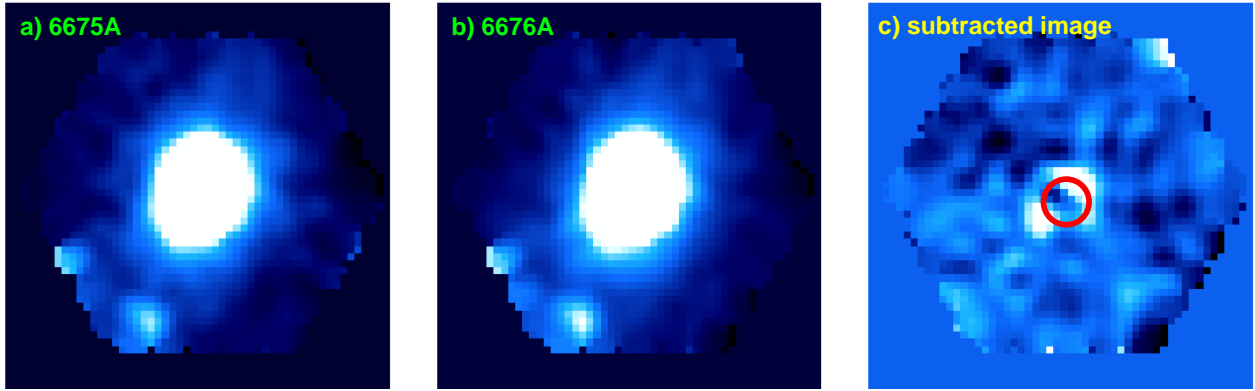}
  \caption{MaNGA data of SDSS J170124.01+372258.0. a) IFU image at 6675 $\AA$. b) IFU image at 6676 $\AA$. c) subtracted image from b) - a). The arc pattern identified by the circular Hough transform is marked in the red circle.}
  \label{fig:manga}
\end{figure*}

As a proof of concept, we used SDSSJ170124.01+372258.0 to demonstrate the feasibility of the approach mentioned in the previous section.   
We acquired its MaNGA IFU spectra from SDSS SkyServer\footnote{https://skyserver.sdss.org/dr13/en/tools/chart/charthome.aspx}, which covers wavelength from 3622 to 10353 $\AA$ and a spectral
resolution of 1$\AA$. We employed the IRAF\footnote{http://iraf.noao.edu}
routine \textit{imslice} to extract 2D IFU images from the 3D IFU
data-cube (see Fig. \ref{fig:manga}).
We then used the IRAF routine \textit{imarith}
to subtract the 2D IFU images. Examples of IFU images at a given wavelength and lensed arcs from subtracted images can be found in
Fig. \ref{fig:manga}. 
We then applied Hough transform algorithm
to the MaNGA 6676$\AA$ - 6675$\AA$ IFU image,
where the [OII] emission from the background source form an arc pattern
(Fig. \ref{fig:manga}). It only takes 0.13 seconds for Hough transform
to identify this
arc pattern on a 2.7 GHz Intel Core i5 processor.

For a single object, it takes 20 seconds to download the MaNGA data,
5 seconds to
extract 2D IFU images from the 3D data-cube, 15 seconds for
image subtraction, and 875 seconds for Hough transform
to analyze all subtracted images.
To summarize, this only costs $\sim$ 15 minutes to fully analyze
one single object.

\section{Prospects}
Using the recently discovered MaNGA lens by \cite{2017MNRAS.464L..46S},
we demonstrate the feasibility of an automated approach to identify strong gravitational lenses in IFU surveys.
This approach is applicable to other IFU surveys,
e.g. CALIFA, SAMI, Hector, DESI, and 4MOST,
which will all deliver IFU data-cube in similar format as MaNGA.
The arc identification procedure can be improved with more
realistic lens modeling, e.g. ARCFINDER \citep{2006astro.ph..6757A} for arcs, or
CHITAH \citep{2015ApJ...807..138C} for multi-imaged quasars, at the cost of computation time.
Nevertheless the Hough transform approach (as proposed here) already provides a simple solution
to detect arc patterns in a timely manner for the big data to come.

\end{CJK*}

\end{document}